\documentclass[prl,letterpaper,twocolumn,final,superscriptaddress,floatfix]{revtex4}

\usepackage{graphicx,psfrag,amsmath,amssymb,amsfonts,bbm,latexsym,color,dcolumn,bm,mathbbol}
\usepackage[framemethod=tikz]{mdframed}
\usepackage{hyperref}
\usepackage{xfrac} 
\usepackage{ifthen}

\usepackage{amsmath,amssymb}             
\usepackage[french, english]{babel}
\newcommand{\refn}[1]{Eq. (\ref{#1})}
\newcommand{\smartit}[3]{\int^{#1}_{#2} \! \mathrm{d} #3 \,}  
\newcommand{\braket}[2]{\bigl\langle#1\bigl|\bigr.#2\bigr\rangle}

\newcommand{\pa}{\partial}

\usepackage{enumerate}

\newcommand{\cverbose}{1}

\begin{document}

\graphicspath{{Figures/}}

\title{Velocity condensation for magnetotactic bacteria}

\author{Jean-Fran\c cois Rupprecht}
\affiliation{Ecole Normale Sup{\'e}rieure, Laboratoire de Physique Statistique, UMR CNRS 8550, 24 rue Lhomond, Paris  (France).}
\affiliation{Mechanobiology Institute, National University of Singapore, Singapore.}

\author{Nicolas Waisbord}
\affiliation{Institut Lumi\`ere Mati\`ere, UMR CNRS 5306, Universit\'e Lyon 1, Lyon (France).}

\author{Christophe Ybert}
\affiliation{Institut Lumi\`ere Mati\`ere, UMR CNRS 5306, Universit\'e Lyon 1, Lyon (France).}

\author{C{\'e}cile Cottin-Bizonne}
\affiliation{Institut Lumi\`ere Mati\`ere, UMR CNRS 5306, Universit\'e Lyon 1, Lyon (France).}

\author{Lyd\'eric Bocquet}
\email{lyderic.bocquet@ens.fr}
\affiliation{Ecole Normale Sup{\'e}rieure, Laboratoire de Physique Statistique, UMR CNRS 8550, 24 rue Lhomond, Paris  (France).}

\begin{abstract} 
Magnetotactic swimmers tend to align along magnetic field lines against stochastic reorientations.  We show that the swimming strategy, e.g. active Brownian motion versus run-and-tumble dynamics, strongly affects the orientation statistics. The latter can exhibit a velocity condensation whereby the alignment probability density diverges. As a consequence, we find that the swimming strategy affects the nature of the phase transition to collective motion, indicating that L\'evy run-and-tumble walks can outperform active Brownian processes as strategies to trigger collective behavior.
\end{abstract}

\date{\today}

\ifthenelse{ \cverbose > 1}{}{\maketitle}

Bacteria, spermatozoa or algae have in common the ability to propel themselves in low-Reynolds fluids in order to explore space \cite{Purcell1977,Cates2012}. 
The directed motion of these swimmers is always affected by stochastic impulses due to noise in the propulsion mechanism. Swimmers undergoing white noise perturbations, which lead to persistent small-amplitude fluctuations of the orientation \cite{Lowen1,Romanczuk2012}, are usually called active Brownian particles (ABPs). In contrast, Bacteria like E. Coli exhibit sudden reorientations of their velocity vector (called tumbles) which are due to stochastic switches in the direction of rotation of propelling flagella \cite{Berg2004}. Such dynamics are usually coined as Run-and-Tumble (RT). 
Though ABPs and RTs correspond to two different swimming strategies, in the absence of external torques, their long-time dynamics are similar and lead in both cases to an effective diffusion process \cite{Cates2012}. 


Biological microswimmers can also orient themselves in response to external stimuli, either of chemical or mechanical nature. In particular, the RT walk is essentially thought to provide a mean to move along chemical gradients, called chemotaxis, in which the run duration is modulated with respect to the direction of the stimulus. Other micro-organisms have also developed the ability to orient their propelling direction under external mechanical fields, for example under gravity (gravitaxis) or shear or flow gradients (gyro- and rheo- taxis) \cite{Stocker1,Hader2005,Fukui1985,Garcia2013}. Similar behavior have been recently reproduced with artificial catalytic swimmers \cite{Palacci2015,Lowen2,Stark,Gao2012}. 

Here we consider the dynamics of swimmers driven under {\it magnetic torques}, keeping in mind that results generalize to a larger class of mechanical torques.  A representative example are magnetotactic bacteria (MB) that behave as self-propelled compasses, due to iron-based organelles orienting the propelling flagella along the magnetic field lines. Since their discovery in 1975, theoretical studies of MB focused on the case of white noise perturbations on the orientation \cite{Blakemore1982,Nadkarni2013}. Recent work also demonstrated how superparamagnetic beads could be attached to E-Coli bacteria, making them reactive to magnetic fields \cite{Dekker2015}. 

Here we show that in the presence of an external aligning field, the orientation distribution strongly differs for the two swimming strategies, ABPs and RTs. For RTs, we report a velocity condensation phenomenon which is associated with a divergence of the orientation distribution function in the direction of the field and which occurs above a critical magnetic field. We point that the resulting behavior is significantly different from a chemotactic response. In the final paragraph, we consider the onset of the collective phase of a swarm with nematic interactions. We show that the nature of the alignment divergence shapes the phase diagram of the isotrope-nematic transition.

The ABPs dynamic is a diffusion process on the direction of the velocity vector $\bm{V}$, with a fixed speed $\lvert \bm{V} \rvert = V_0$ \cite{Lowen1,Romanczuk2012,Blakemore1982}; hence the dynamics of the alignment angle $\theta$ is described by an Ito equation: 
$ d\theta = f(\theta) dt/ \tau_B + \sqrt{2 dt D_r}\, \zeta$, where $\zeta$ is
a Gaussian white noise with $\left\langle \xi(t) \xi(t^{\prime}) \right\rangle = \delta_{t,t'}$ and $D_r$ is a rotational diffusion coefficient, and $f(\theta)= -\sin(\theta)$ is the magnetic torque. The magnetic relaxation time $\tau_B$ can be expressed as $\tau_B = \xi_0/(m B_a)$, where $m$ is the magnetic moment and $\xi_0$ is a rotational drag coefficient. The stationary probability distribution for $\theta$ corresponds to the Boltzmann statistics 
\begin{align} \label{eq:abp}
P_{\infty}(\theta) = \mu(\theta) \exp\left[ 1/(\tau_B D_r) \cos \theta\right]/Z_d,
\end{align}
where $Z_d$ a dimension dependent normalization factor and $\mu(\theta)$ is the uniform probability measure ($\mu(\theta)=1/\pi$ in 2D or $\sin(\theta)/2$ in 3D). 
In 3D, the mean velocity $V_z = \left\langle \cos \theta \right\rangle $ reduces to
$V_z = V_0 f(1/(\tau_B D_r))$, where  $f(x) = \coth x - x^{-1}$ is the classical Langevin function  \cite{Blakemore1982}.  Indeed, \refn{eq:abp} corresponds to the distribution of a passive magnet in a thermal noise \cite{Nadkarni2013}, with an effective temperature defined as $k_{B} T_{\rm eff}=D_r\xi_0$. 

\begin{figure}[t]
\ifthenelse{ \cverbose > 2}{}
{	\includegraphics[width=8cm]{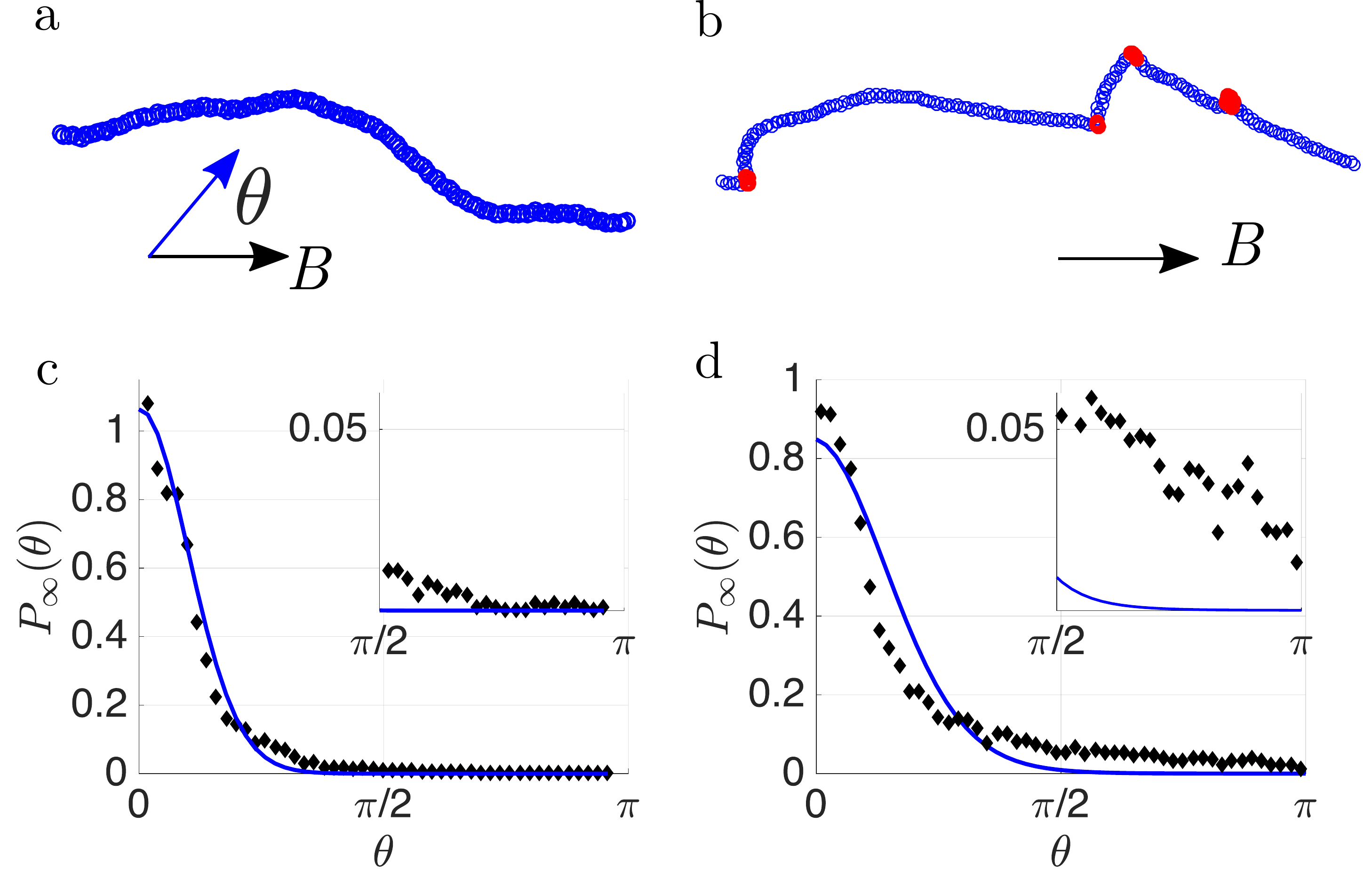} 

	}
	\caption{(color online) Trajectories of MB in (a) rich growing medium and (b) poor growing medium environments. (c--d) Alignment angle distribution at $B_a = 7 \cdot 10^{-5} \, \mathrm{T}$ in (c) rich growing medium and (d) poor growing medium environments:  (black crosses)  experimental histogram and  (solid blue line) best ABP fit with (a) $B/D_\perp = 11$ and (b) $B/D_\perp = 4.5$. In (d), there is a $10\%$ discrepancy between the experimental and fitted cumulative distributions (i.e. Kolmogorov-Smirnov norm \cite{Massey1951}). (Insets) The anti-parallel response is enhanced in a poor environment (see \cite{arxiv}). }
	\label{fig:1}
\end{figure}

From an experimental point of view, MB bacteria usually behave as ABP particles in  standard chemical environment (see \cite{Frankel1979}, \cite{arxiv} and Fig. \ref{fig:1}a. c. below). However, it has been recently reported that some specific environments can trigger non-Brownian reorientations of MB, as demonstrated on the strains MO-1 \cite{Zhang2014a} and MC-1 \cite{arxiv}. In particular, it has been shown in \cite{arxiv} that the trajectory of the bacteria presents sudden changes of direction when the concentration in growing medium is reduced (see Fig. \ref{fig:1}b). These kinks are then detected by a tracking algorithm, which show that the run durations are exponentially distributed (see \cite{arxiv}). 
Using the experimental data reported in \cite{arxiv}, we build 
the experimental histogram presented in Fig. \ref{fig:1}d from 500 trajectories. The histogram is peaked in the direction of the magnetic field ($\theta = 0$) while maintaining a substantial statistical weight for the anti-parallel
orientation ($\theta = \pi$). These two features can {\it not} be
consistently accounted by an ABP model, which fails to match the
statistical weights both in the parallel and in the anti-parallel directions of
the magnetic field, as highlighted see in Fig. \ref{fig:1}d). This
inconsistency of the ABP model to reproduce experimental results calls for a shift from the classical
Langevin paradigm \cite{Blakemore1982}, which can only describe accurately the behavior of magnetotactic bacteria in a medium favorable to growth. As indicated in \cite{arxiv}, this change in the behavior of magnetotactic bacteria in a lesser favorable environment could be related to an evolutionary advantage. 

{\it RT walk --} The RT dynamics is composed of runs at a fixed speed $V_0$ interrupted by instantaneous reorientations. During runs under a magnetic field $\bm{B_a}$, the evolution of the alignment angle $\theta$ (between $\bm{V}$ and the applied magnetic field $\bm{B_a}$) is deterministic, with  $ \dot{\theta}= f(\theta)/\tau_B$.  Furthermore, we assume that the duration of each run, $x$, (i) is drawn according to a given probability density $\rho(x/\tau_r)$, (ii) is independent of the previous runs (iii) is independent of the alignment direction $\theta$ (in contrast to chemotaxis).
After a tumble, the alignment angle $\theta_0$ is drawn according to the uniform probability measure $\mu(\theta_0)$.
We finally define the magnetotactic dimensionless parameter $B$ as
\ifthenelse{ \cverbose > 1}{}{
\begin{align} \label{def:magnetotactic_constant}
B  = \tau_r\frac{m \vert B_a\vert}{\xi_0},
\end{align}}
where $\tau_r$ is the mean run time, so that  $B  = \tau_r/\tau_B$. Remarkably, the estimated values for the magnetotactic number $B$ appear to be of the order unity for MB in typical geomagnetic fields (see \cite{arxiv} and \cite{Nadkarni2013}).

{\it Angular distribution and velocity condensation --}. We seek to obtain the expression for the stationary probability $P_{\infty}(\theta)$ density for the RT walk, defined so that the probability that the angle $\theta$ belongs to the interval $ \left[\theta_1, \theta_2\right]$ reads $\int^{\theta_2}_{\theta_1} \! P_{\infty}(\theta) \mathrm{d} \theta$. Partitioning on successive events, we find that $P_{\infty}(\theta) = \int^{\infty}_{0} \! \mathrm{d}t  \, \pi(t) \int^{\pi}_{0} \! \mathrm{d}\theta_0 \, \mu(\theta_0)\,  \delta(\theta - \theta_t(\theta_0))$,
where $\pi(t)$ is the distribution of 'run' time since the last tumble, and which is to be determined from the distribution of run duration $\rho(x)$ by renewal process theory \cite{Feller1968}; $\theta_0$ is the  outgoing angle after the tumble; and $\theta_t(\theta_0)$ is the time-dependent evolution operator. For a torque with angular dependence $f(\theta)$, the latter is formally  defined as  $\theta_t(\theta_0) = F^{(-1)}[F[\theta_0]+B t] $, with $F$ a primitive of $1/f$ and $F^{(-1)}$ the reciprocal function of $F$. Using that the function $\theta_0 \rightarrow \theta-\theta_t[\theta_0]$ is canceled for $\theta_0 = \theta^{*}_0(\theta, t) = F^{(-1)}[F[\theta]-B t]$, one gets
\ifthenelse{ \cverbose > 1}{}{
\begin{align} \label{eq:main}
 P_{\infty}(\theta) f(\theta) &= \int^{\infty}_{0} \! \mathrm{d}t \, \pi(t)  (\mu.f)[\theta^{*}_0(\theta, t)].
\end{align}}
Equation (\ref{eq:main}) holds for an arbitrary torque $f(\theta)$ and spent time distribution $\pi(t)$.

\begin{figure}[t!]
\ifthenelse{ \cverbose > 2}{}
{	
\includegraphics[width=8.5cm]{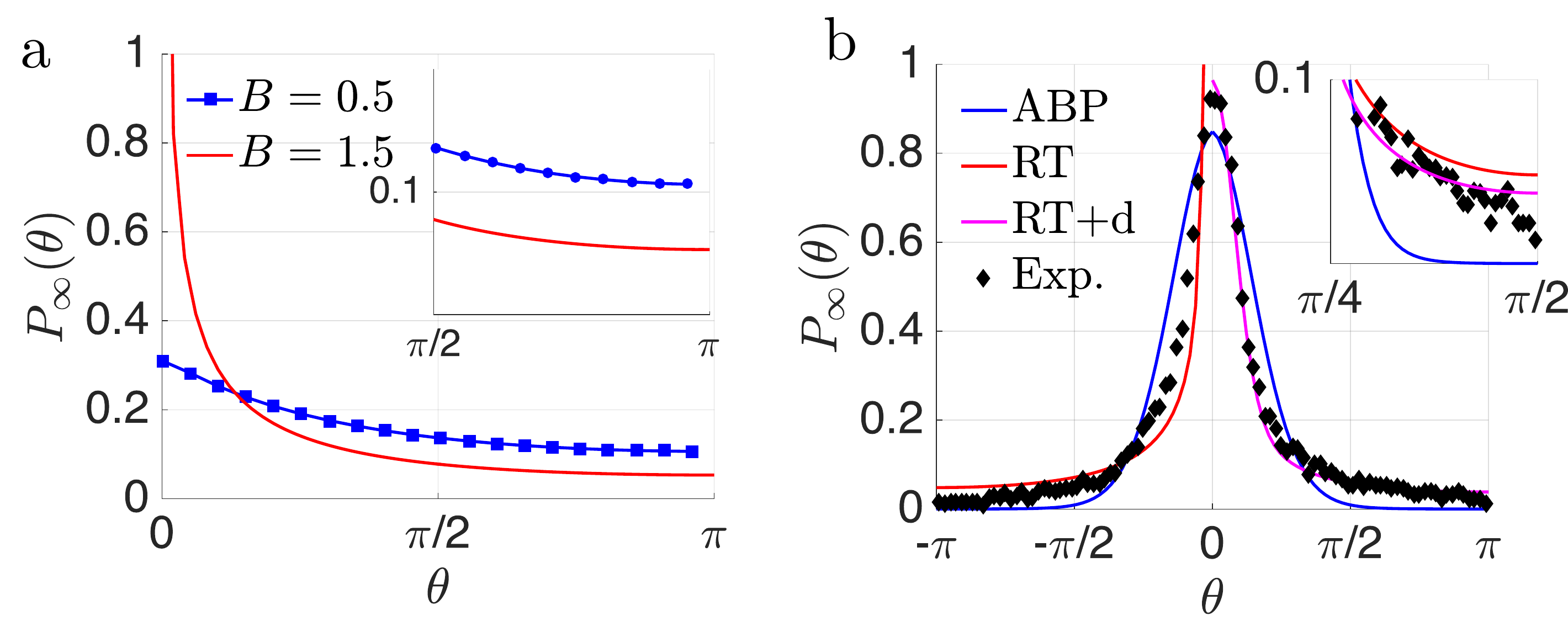}}
	\caption{(color online) (a) Probability density $P_{\infty}(\theta)$ for RTs (2D) for (red solid line) $B=4$ and (blue circle line) $B=0.25$. (Inset) The distribution in the opposite direction to the magnetic field remains comparable for both $B=0.25$ and $B=4$. (b) Fit of the experiments by the RT model: (black crosses) experimental histogram; (positive side, red solid line) RT fit with $B=2.3$; (negative side, magenta solid line) RT fit with a mild rotary diffusion with $D^{\prime}_r/\tau_r = 0.15$ and $B=3.2$; (both sides, blue solid line) ABP fit with $B/D_\perp = 4.5$. The Kolmogorov-Smirnov error is of $7\%$ for the pure RT model and of $3\%$ for the perturbed RT model. }
	\label{fig:2}
\end{figure}

We first consider a magnetic torque $f(\theta) =- \sin(\theta)$ and  $\rho(t) = \exp(-t)$ (exponential RT). The spent time distribution is then $\pi(t) =\rho(t)$ \cite{Feller1968}. Following the previous definition, we obtain $\theta_0^{*} = 2 \arctan[\tan (\theta/2) \exp({B t})]$.
In 2D, 
we apply \refn{eq:main} and we obtain 
\ifthenelse{ \cverbose > 1}{}{
\begin{align}\label{eq:2D_pin}
 P_{\infty}(\theta) &= \frac{1}{B} {\left(\tan{\theta/ 2}\right)^{\sfrac{1}{B}} \over \sin\theta} 
\smartit{\pi}{\theta}{\phi} {\mu(\phi) \over \left(\tan{\phi/2}\right)^{\sfrac{1}{B}}}
\end{align}}
A key feature which emerges from Eq. (\ref{eq:2D_pin}) is that the low-$\theta$ behavior drastically differs above and below the critical value $B_c=1$ (see Fig. \ref{fig:2}a). For $ B < B_c$,  $P_\infty(\theta)$ takes a finite value 
at $\theta = 0$. 
However, for $B > 1$ we find that
\ifthenelse{ \cverbose > 1}{}{
\begin{align} \label{eq:2D_pinf}
 P_{\infty}(\theta) \underset{\theta \rightarrow 0}{\sim} \gamma^{-1}_d \theta^{-(1 - \sfrac{1}{B})},
\end{align}}
where $\gamma_2 = 2^{\sfrac{1}{B}} B \cos \left(\pi/(2 B)\right)$ in 2D and  $\gamma_3 = 2^{\sfrac{1}{B}+1} B^2 \sin(\pi/(2 B))/\pi$ in 3D. At $B = B_c$  a dynamical transition occurs above which the probability density diverges 
-- a property that we call the velocity condensation. 

{\it Extension and robustness --}
We first remark that the condensation phenomenon is maintained for alternative aligning torques, provided the torque is strong enough around $\theta = 0$. Consider that $f(\theta) \sim -\theta^n$ for $\theta \ll 0$, then: (i) if $n > 1$ and for exponential RT, the torque term is too weak for the velocity condensation to occur, and (ii) if $n < 1$, the condensation always occurs, as the $\theta = 0$ state can be attained after a finite run time. Second, the strict mathematical divergence disappears in the presence of a rotary Brownian noise on the velocity orientation during runs, characterized by a diffusion coefficient $D^{\prime}_r$. As $\theta \rightarrow 0$, the diffusive noise eventually dominates over the vanishing torque and the orientation probability scales as $P_{\infty}(\theta) \propto \exp(-\theta^2 B \tau_r/(2D^{\prime}_r))$ in the region $\theta \in \left[0, D^{\prime}_r/(\tau_r B)\right]$). 
However, provided that the rotary diffusion coefficient noise is relatively small ($D^{\prime}_r/\tau_r < 1$), the probability density of RTs is sharply peaked when $B >1$.

{\it RT fit of experiments --} We now compare the prediction of the RT model to the experimental histogram presented in \ref{fig:2}b.  In contrast to the ABP model fit, which can not account for the sharp peak in the orientation distribution without underestimating it in the anti-parallel directions, the RT walk provides the appropriate statistical weight to both the parallel and the anti-parallel directions. The increase in the quality of the fit can be measured through the Kolmogorov-Smirnov norm \cite{Massey1951}, which quantifies the discrepancy between the cumulative distributions. The quality of the fit is increased by considering a RT walk in which runs are perturbed by a mild rotary noise ($D^{\prime}_r/\tau_r = 0.15$). We conclude that, in spite of this small rotary diffusion that affects the orientation of bacteria, the distribution in Fig. \ref{fig:2} is still very sharply peaked in the direction of the magnetic field.

We conclude that the RT walk is more efficient than the ABP process to sample both the parallel and anti-parallel directions to the magnetic field. Our intuitive explanation is that, in contrast to a diffusion process, all orientations are sampled after a tumble, and in particular the anti-parallel directions to the magnetic field.  

{\it Mean velocity and diffusion --}
From the orientation distribution, we can calculate the mean velocity of the RTs. The averaged velocity in the direction of the magnetic field is defined as $V_z = V_0 \left\langle \cos \theta \right\rangle$.
Using the previous expressions for the distribution function, one gets
\ifthenelse{ \cverbose > 1}{}{
\begin{align} \label{eq_vz}
V_z=V_0 \times \frac{1}{B} \int_0^1 dw\, w^{\sfrac{1}{B}-1} g_d(w).
\end{align}}
In 2D, $g_2(w) = (1-w)/(1+w)$ and $V_z/V_0  =  \left\lbrace \psi ^{(0)}\left(\frac{B+1}{2 B}\right) - \psi ^{(0)}\left(\frac{1}{2 B}\right) \right\rbrace/B-1$, with $\psi^{0}(z)= \Gamma^\prime(z)/\Gamma(z)$ and $\Gamma(z)$ is the Gamma function \cite{Ryzhik1957}. 
In 3D, $g_3(w) = \left(1-w^4+4 w^2 \log (w)\right)/\left(w^2-1\right)^2$ and \refn{eq_vz} reads
$V_z/V_0  = \psi^{(1)}\left( \sfrac{1}{(2 B)}\right)/(2 B^2) - 1/B -1$,
where $\psi^{1}(z)$ stands for the derivative of $\psi^{0}(z)$ \cite{Ryzhik1957}.
This result is plotted in Fig.~2 against the results obtained for ABP in terms of the Langevin equation.
Interestingly, there is no strong signature of the onset of the velocity condensation on the mean velocity. Furthermore, while the expression for $V_z$ differs from the Langevin prediction for ABPs, we observe that 
for a general $B$, the curve of the RT velocity lies in between those for ABPs with $D_r = 1/(2\tau_r)$ and $D_r = 1/(\tau_r)$ (see Fig. \ref{fig:3}a). We obtain similar conclusions concerning the diffusion coefficient $D_{\perp}$ in the transverse direction $(xOy)$ (see \cite{SI}), suggesting that, in spite of the velocity condensation, RTs are  as efficient as ABPs in exploring their environment.

\begin{figure}[t!]
\ifthenelse{ \cverbose > 2}{}
{	
\includegraphics[width=8.3cm]{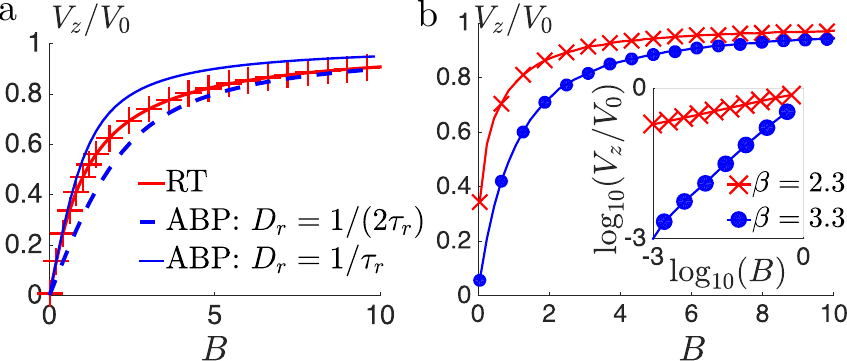}}
	\caption{(color online) Forward mean velocity $V_z$ in 3D, normalized by the velocity norm $V_0$: (a) (red line)  exact expression for the velocity of a RT with an average step duration $\tau_r$; (blue solid line) ABP with $D_r = 1/\tau_r$; (blue dashed line) ABP with $D_r = 1/(2 \tau_r)$. (b) RT L\'evy walks case ($\rho(x) \sim x^{-\beta}$, $\beta \leq 3$): (red crosses) $\beta = 2.30$ and (blue circles) $\beta = 3.30$. Inset: log-scale behavior at $\theta \ll 1$. Inset: in the limit $B \ll 1$, $V_z$ scales as $B^{\xi}$ (black line), with $\xi=\beta-2=0.30$ for $\beta=2.30 < 3$ and $\xi=1$ for $\beta = 3.30 \geq 3$.}
	\label{fig:3}
\end{figure}

\paragraph{Chemotaxis --}
Chemotaxis refers to an angular dependence in the mean duration of a run \cite{Berg2004}. We first consider the parallel chemotaxis case $\tau_r(\theta) = 1 + \chi \cos(\theta)$, which favors runs in the direction of the magnetic field when $\chi >0$. In units of $\tau_r$, the Fokker-Planck equation reads \cite{Bearon2000}
\begin{align*} 
B \pa_\theta (\sin \theta P) - (1+\chi \sin \theta)  P(\theta) = -\left( 1 + \chi \left\langle P \sin \theta \right\rangle \right),
\end{align*}
where $ \braket{P}{\sin \theta}  = \smartit{\pi}{-\pi}{u} P(\phi) \sin(\phi)$. We show that the solution reads
\begin{align}  \label{eq:solution_chemotaxis}
P(\theta)=  \gamma \frac{\tan(\theta/2)^{\sfrac{1}{B}}}{\sin(\theta)^{1-(\sfrac{\chi}{B})}} \smartit{\pi}{\theta}{\phi}
\frac{\sin(\phi)^{-\sfrac{\chi}{B}}}{\tan(\phi/2)^{\sfrac{1}{B}}}, 
\end{align}
for $\theta >0$, where $\gamma = 1 + \chi \braket{p}{\sin \theta} $. The constant $\braket{p}{\sin \theta}$ is found as a solution of the self-consistency equation: $\braket{p}{\sin \theta}  = \braket{f_p}{c}/(\lambda_0 - \epsilon \braket{f_p}{c})$. From \refn{eq:solution_chemotaxis}, we find that a positive parallel chemotaxis lowers the value of the critical magnetotatic constant above which the velocity condensation occurs, as $B_c = 1-\chi$. In contrast, a transverse chemotactic field, as defined by $\tau_r(\theta) = 1 + \chi \sin(\theta)$, will not change the value $B_c = 1$ (see SI).

{\it L\'evy walks --}
We show that the velocity condensation phenomenon is further amplified for systems exhibiting a L\'evy statistics of the run period. 
L\'evy walks are characterized by heavy--tailed distribution of run duration: $\rho(x)  = \mathbb{1}_{x>1} (\beta-1)/x^{\beta}$ with $2 < \beta < 3$.  The value $\beta=2$ corresponds to a predicted optimal search strategy \cite{Rosa1999}. To apply \refn{eq:main}, we notice that $\pi(t)  =(\beta-2)/(\beta-1) t^{1-\beta}$, when $t>1$ \cite{Feller1968}, and that the function $t \rightarrow \sin(\theta_0^{\star})$ is sharply peaked around $t^{*} = -\log(\tan(\theta/2))/B$. We find that the velocity condensation occurs for any $\beta>2$ as
\ifthenelse{ \cverbose > 1}{}{
\begin{align} \label{eq:asymptoticlevy}
P_{\infty}(\theta)\underset{\theta \rightarrow 0}{\sim} \gamma \frac{B^{\beta-1}  (\beta -2)}{ (\beta -1)} \frac{1}{\theta (\log (1/\theta ))^{\beta-1}}, 
\end{align}}
where $ \gamma  = 0.46 \ldots$ both in 2D and 3D. This expression corresponds to an enhanced condensation compared to exponentially-distributed runs. The mean velocity is found in terms of an expression analogous to \refn{eq_vz}. We truncate the functions $g_d(w)$ to its first order expansion at $w=1$ to obtain:
\ifthenelse{ \cverbose > 1}{}{
\begin{align} \label{eq:asymptoticvelocitylevy}
V_z/V_0  \underset{B \rightarrow 0}{\sim} \gamma_d \frac{ \Gamma (3-\beta )}{\beta -1} B^{\beta-2},
\end{align}}
where $\gamma_2 = 1/2$ and $\gamma_3 = 2/3$.
The non-analytical scaling $B^{\beta-2}$ in \refn{eq:asymptoticvelocitylevy} corresponds to a highly sensitive directional response at the onset of the stimulus detection (from $B = 0$ to $B > 0$). In comparison, the velocity $V_z$ is proportional to $B$ when $B \ll 1$ for ABPs as well as for RTs with a finite second moment for the run duration (e.g. $\beta \geq 3$ in Fig. \ref{fig:3}b). 

{\it Collective behavior --} We finally consider the consequence of the velocity condensation on the collective behavior of a swarm \cite{DoiEdwardsBook,Bolley2012, Degond2011, Ezhilan2013,Bertin2015,Peruani2012,Baskaran2008, Baskaran2008a}. We adapt the Maier-Saupe mean-field treatment for a highly concentrated swarm of interacting self-propelled rods \cite{DoiEdwardsBook} (see also \cite{Bolley2012, Degond2011, Ezhilan2013}). In contrast to previous results which assumed a Boltzmann distribution for the orientation distribution \refn{eq:abp}, we use here the precise statistic of the orientation from \refn{eq:main}. Following  \cite{DoiEdwardsBook, Ezhilan2013}, we consider that interactions between bacteria result in an effective torque acting uniformly on each bacterium: $f(\theta) = -  U_0 A[P_\infty] \sin(2 \theta)$, where $U_0$ is the interaction strength and $A$ measures of the local nematic order and is defined as $A[P_\infty] = \smartit{\pi}{0}{u} \cos(2 u) P_\infty(u)$
in 2D and $A[P_\infty] = \smartit{\pi}{0}{u} (3 \cos^2 u - 1) P_\infty(u)$  in 3D. Using \refn{eq:main}, we compute the probability distribution $P_\infty(S)$ that corresponds to an imposed value $S = A[P_\infty]$. The order parameter, denoted $S^{\star}$, is then found as the solution of the following self-consistency equation: $S^{\star} = A[P_\infty(S^{\star})]$. 
For RTs with exponential runs, the isotropic phase ($S^{\star} =0$) is destabilized above a critical value of the interaction strength $U^{(c)}_0 > 1.87. \tau^{-1}_r$ in favor of the nematic phase. The phase diagram is alike for ABP swimmers \cite{Ezhilan2013}. However, the probability distribution diverges in both directions $\theta = 0$ and $\theta = \pi$ for RTs within the nematic phase (3D). Within Onsager's theory, the quantity $1/U^{(c)}_0$ can be interpreted as an excluded volume induced by steric interactions \cite{Onsager1949}. 

\begin{figure}[t!]
\ifthenelse{ \cverbose > 2}{}
{	
\includegraphics[height=3.2cm]{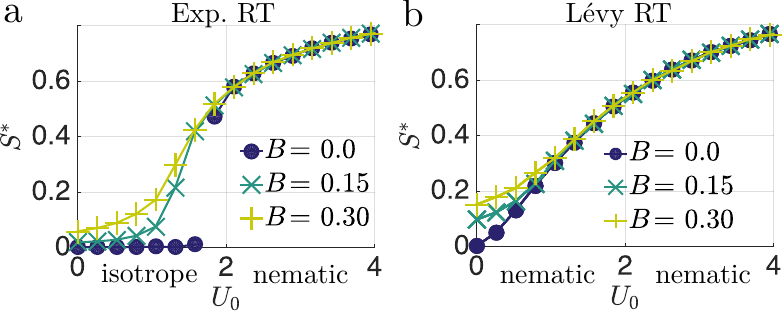}}
	\caption{(color online) Phase diagrams of the order parameter $S^{*}$ in 3D in terms of the interaction strength $U_0$, and for several values of $B$ (see \refn{def:magnetotactic_constant}): (a) RTs with exponentially-distributed runs, (b) L{\'e}vy walks with $\beta = 2.6$: $S^{*} > 0$ for $U_0 > 0$ even at $B = 0$ (blue circle curve), hence the isotropic phase is intrinsically unstable.}
	\label{fig:4}
\end{figure}

For RT L\'evy walks, the phase diagram is drastically changed due to destabilization of the isotrope phase (see  Fig. \ref{fig:4}). Indeed, the order parameter should satisfy by the following self-consistency equation:
\ifthenelse{ \cverbose > 1}{}{
\begin{align} \label{eq:levycollective}
S^{\star}   = \gamma_d \frac{ \Gamma (3-\beta )}{\beta -1}  (U_0 S^{\star})^{\beta - 2}, \quad S^{\star} \ll 1,
\end{align}}
where $\gamma_3 = (2^\beta - 4)/5$ in 3D. Due to the behavior of the probability distribution in \refn{eq:levycollective}, \refn{eq:asymptoticvelocitylevy} displays an non-analytical behavior with $S^{\star} \ll 1$. This $S^{\beta - 2}$ behavior implies that  a solution $S^{\star} > 0$ necessarily exists for any value of the interaction strength $U_0 > 0$. Hence, we find that the isotropic phase is intrinsically unstable, as $U^{(c)}_0 = 0$, which corresponds to a diverging excluded volume. Similar conclusions can be drawn in 2D ($\gamma_2 = 1/2$ in \refn{eq:levycollective}).

Experimentally, it appears that bacteria \cite{Korobkova2004,Ariel2015} and immune cells \cite{Harris:2012fk} can perform L\'evy walks with an exponent $\beta < 3$, which is within our predicted regime of a high sensitivity at the onset of the stimulus detection and to collective motion (see Eqs. \ref{eq:asymptoticvelocitylevy} and \ref{eq:levycollective}). 


{ \it Conclusion -- } In this paper, we exhibit a divergence in the orientation response of the RT walk under torque. This divergence is required to account the high directional response of tumbling magnetotactic swimmer, even when perturbed by a Brownian rotary noise (see \ref{fig:2}b). Experiments on MC--1 bacteria confirm the observation that tumbling bacteria exhibit a stronger parallel or anti-parallel response to the magnetic field, which cannot be described by the standard ABP model. Based on our analytical expressions for the orientation distribution, we find that the noise statistic has a crucial impact on the onset of collective motion. The fact that for L\'{e}vy runs, the transition occurs in the limit of an infinite excluded volume hints at a possible extension of Onsager's theory \cite{Onsager1949} in terms of a dynamical excluded volume that depends on the noise statistic.


\ifthenelse{ \cverbose > 1}{}{
\acknowledgements{We thank J. Prost for suggesting the idea of a dynamical excluded volume. We also thank Fran\c cois Detcheverry for fruitful discussions. NW was supported by the AXA fond.}
}

\newpage
\onecolumngrid
\appendix

\section*{Supplemental Material}

We recall that $P_{\infty}(\theta) = \int^{\infty}_{0} \! \mathrm{d}t  \, \pi(t) \int^{\pi}_{0} \! \mathrm{d}\theta_0 \, \mu(\theta_0)\,  \delta(\theta - \theta_t(\theta_0))$, which corresponds to Eq. (2) in the main text. The argument of the delta function is canceled for $\theta_0 = \theta^{*}_0(\theta, t) = F^{(-1)}[F[\theta]-B t]$,  with $F$ a primitive of $1/f$ and $F^{(-1)}$ the reciprocal function of $F$. We finally obtain:
\begin{align}\label{si:general}
 P_{\infty}(\theta) &= {1 \over \sin \theta} \int^{\infty}_{0} \! \mathrm{d}t \, \pi(t) \sin(\theta^{*}_0(\theta, t)).
\end{align}
We recall the identity 
\begin{align} \label{eq:fondamentale}
\sin(\theta_0^{*}) = \frac{2 e^{B t} \tan \left(\theta/2\right)}{1 + e^{2 B t} \tan ^2\left(\theta/2\right)}.
\end{align}

The change of variable $B t = F(\theta)$ leads to 
\begin{align}
 P_{\infty}(\theta) &= - {1 \over B f(\theta)} \int^{\pi}_{\theta} \!  \! \mu(\phi) \mathrm{d}\phi \ \pi\left(\left\lbrace F[\theta] - F[\phi] \right\rbrace/B\right).  \label{eq:main2}
\end{align}
The expression \refn{eq:main2} lead to the identities of Eqs. (5) and (6) in the main text.

\section{Exponentially distributed runs}

\subsection{Approximate expressions for the probability distribution}

Let us define the function $\phi(v)={e^{-v}} {e^{2 B v} / \left(1+u^2\, e^{2 B v}\right)^2 }$, with $u = \tan(\theta/2)$,
which corresponds to the integrand in \refn{si:general}. We first notice that the function $\phi$ exhibits a maximum for a positive $v=v^\star$ for $B >1/2$ and $u^2<(2 B +1)/(2 B - 1)$. Otherwise the function $\phi$ decays smoothly to zero. Hence we distinguish two cases that will leads to two different approximation schemes.

\subsubsection{Large $B$: $B > 1/2$ and $u^2<(2 B +1)/(2 B - 1)$}

For $B \rightarrow \infty$, the function is peaked around this maximum at $v^\star$.
One calculates $\exp(2 B v^\star)= (2 B-1)/(2 B + 1) \times 1/u^2$ and $v^\star = {1/(2 B)} \times \log[(2 B - 1)/(2 B +1) 1/u^2]$.

We write $\phi(v)= \exp[S[v]]$ and we expand $S(v)$ around its maximum at $v^\star$:
\begin{equation}
	S(v)=\log[\phi(v^\star)] - {1\over 2 \sigma^2} (v-v^\star)^2 + \mathcal{O}((v-v^\star)^3)
\end{equation}
with ${1/\sigma^2} = - {d^2 \over dv^2} [ \log[\phi(v)]]_{\lvert v=v^\star}  = (4 B^2-1)/2$.
The integral over $v$ yields
\begin{align}
\int_0^\infty dv\, \phi(v) \simeq \phi(v^\star) \int_{-\infty}^{+\infty} dv\, e^{-{1\over 2 \sigma^2} (v-v^{\star})^2}  =\phi(v^\star)\sqrt{2\pi \sigma^2}, \label{eq:fphistar}
\end{align}
where
\begin{equation} \label{eq:phivstar}
	\phi(v^\star)={\left({4 B^2- 1 \over B^2 }\right) \left({2 B + 1 \over 2 B - 1}\right)^{1/(2 B)}\times {1\over 16 (\tan \theta)^{2-\sfrac{1}{B}}}},
\end{equation}
From \refn{eq:fphistar}, we obtain the following approximate expression for the distribution $P_\infty(\theta)$, which is expected to be exact in the limit $B \gg 1$ and $u <1$,
\begin{align} \label{eq:pinf_lowalpha_3D}
		P_\infty(\theta) \simeq \frac{\sqrt{\pi } \sqrt{4 B^2-1} \left(1-\frac{2}{2 B+1}\right)^{-\frac{1}{2 B}}}{4 B^2}
		\frac{\tan^{\sfrac{1}{B}}\left(\frac{\theta }{2}\right)}{\sin(\theta)}
\end{align}
with $u=\tan({\theta/2}$. As visible on Fig. \ref{fig:si:1}, \refn{eq:pinf_largealpha_3D} is efficient for $B \gg 1$ and $u <1$. For $B \gg 1$, \refn{eq:pinf_lowalpha_3D} has the same behavior as the expression from Eq. (5) in the main text:
$P_\infty(\theta) \sim L \theta^{-1}$ with a prefactor $L = \sqrt{\pi}/2 = 1.13$ which is comparable to the exact value $L = 1$ presented in the main text (see inset of Fig \ref{fig:si:1}).

\begin{figure}[t]
	\includegraphics[width=5.50cm]{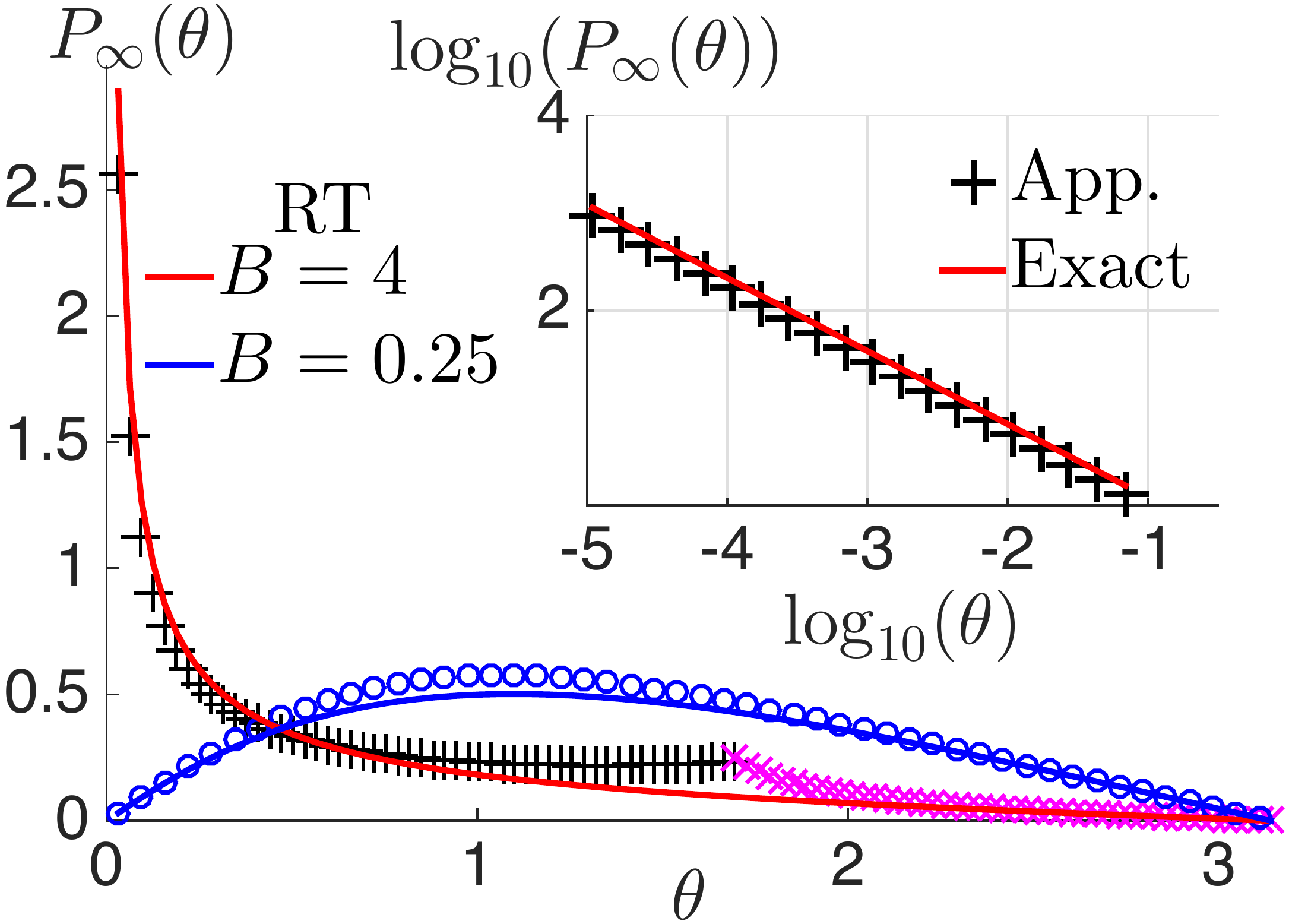} \hskip1cm
	\includegraphics[width=5.50cm]{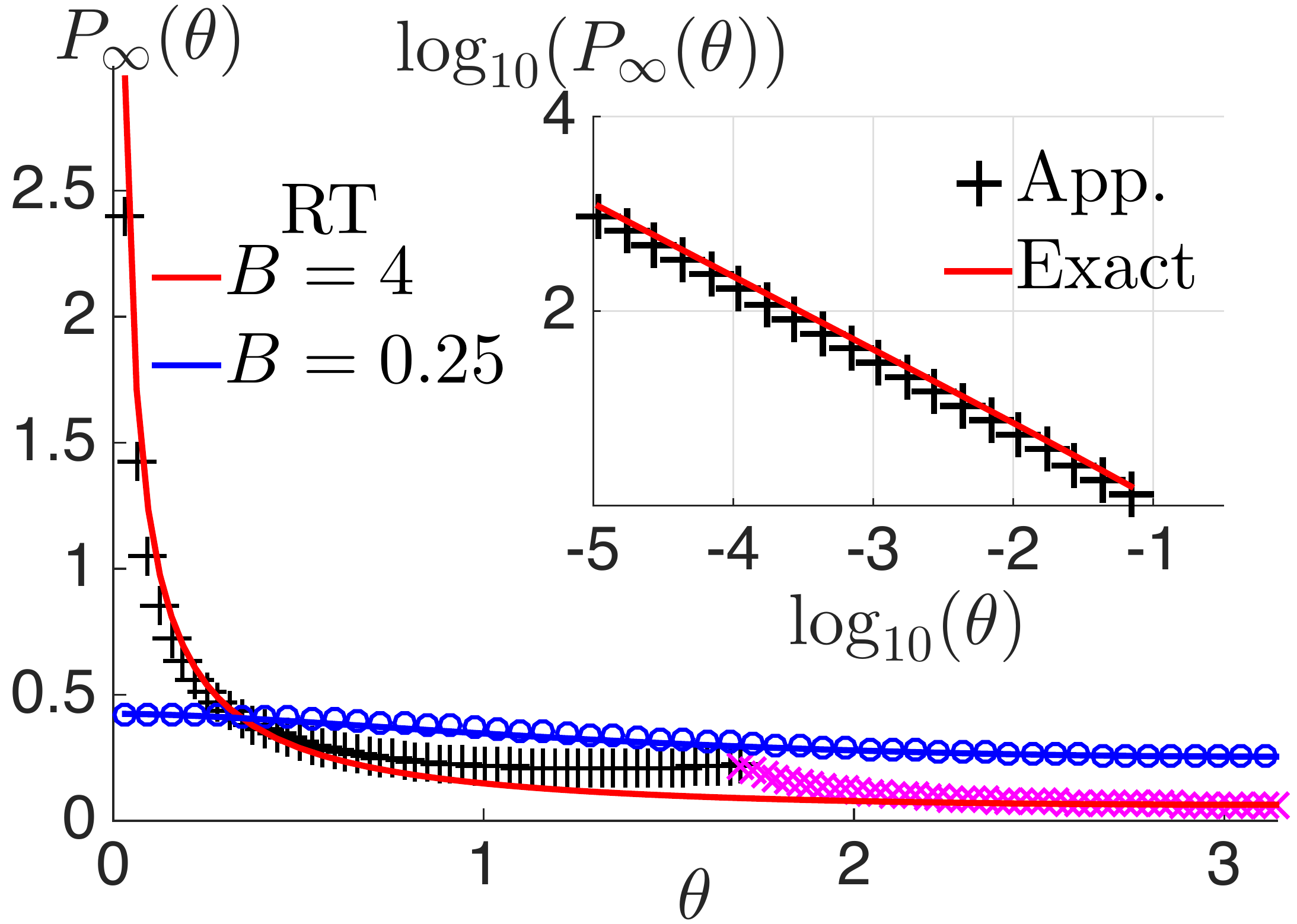}
	\caption{(color online) Stationary probability for the alignment angle $\theta$, for $B = 4$ (red solid line) and $B=0.25$ (blue solid line): (left) 3D case, with (black crosses) the high--$B$ approximate expression from \refn{eq:pinf_lowalpha_3D} and (magenta crosses for $B=4$, blue circles for $B=0.25$) the low--$B$  approximate expression from \refn{eq:pinf_largealpha_3D} (right) 2D case, with (black crosses)  the high--$B$ approximate expression from \refn{eq:pinf_lowalpha_2D} and (magenta crosses for $B=4$, blue circles for $B=0.25$) the low--$B$ approximate expression from \refn{eq:pinf_largealpha_2D}}
	\label{fig:si:1}
\end{figure}

\subsubsection{Small $B$:  $B < 1/2$ or \{$B > 1/2$ and $u^2>(2B+1)/(2B-1)$\}}
In the case $B < 1/2$ or \{$B > 1/2$ and $u^2>(2B+1)/(2B-1)$\}, the function $\phi(v)$ decays smoothly to zero -- exponentially for a large $v$. We make the simplifying substitution that 
\begin{equation}
	\phi(v) \approx \phi(0) \exp[-\gamma t],
\end{equation}
with $\gamma = - \phi^\prime(0)/\phi(0)$. Under this assumption, $\int_0^\infty dv\, \phi(v) \simeq \phi(0)^2/\vert \phi^\prime(0)\vert $, and we obtain that:
\begin{equation} \label{eq:pinf_largealpha_3D}
	{
		P_\infty(\theta) \simeq \frac{\sin \theta}{2} { 1 \over 1- 2 B \cos \theta}.
	}
\end{equation}
As visible on Fig. \ref{fig:si:1}, the approximate expression \refn{eq:pinf_largealpha_3D} works best for $B \ll 1/2$ and $\theta \rightarrow \pi$. 

\subsubsection{Equations in 2D}

In 2D, the analogous equation to \refn{eq:pinf_lowalpha_3D} is for low $B \gg 1$
\begin{align} \label{eq:pinf_lowalpha_2D}
	P_\infty(\theta)  \simeq \frac{1}{\pi} \frac{1+B}{B}  \sqrt{\frac{B^2}{2 (B^2-1)}} \left(\tan ^2\left(\frac{\theta }{2}\right)+1\right) \left(\frac{B-1 }{(B + 1) \tan ^2\left(\frac{\theta }{2}\right)}\right)^{\frac{B-1}{2 B}},
\end{align}
and for $B \ll 1$, the analogous equation to \refn{eq:pinf_largealpha_3D} is
\begin{align} \label{eq:pinf_largealpha_2D}
 	P_\infty(\theta) \simeq \frac{1}{\pi\left(1-B \cos \theta\right)}.
\end{align}
As visible on Fig. \ref{fig:si:1}, the approximate expression \refn{eq:pinf_largealpha_2D} works best for $B\gg 2$ and and $\theta \rightarrow \pi$. 

\subsection{Projected angle} \label{sm:sec:proj}
 
 In tracking experiments, the accessible information is often limited to a projection of trajectories within the 2D focal plane. The observed alignment angle $\psi$ in the focal plane is related to the alignment angle $\theta$ by the relation $\tan(\psi) = \tan(\theta) \sin(\phi)$, where $\phi$ is the azimutal direction (see Fig. \ref{fig:si:1}). The probability distribution $P_{\infty}(\psi)$  for $\psi$ is
 \begin{align}
P_{\infty}(\psi)  = \int^{2\pi}_{0} \! d\phi \, \int^{\pi}_{0} \! d\theta  \, P_{\infty}(\theta) \delta(\psi - \arctan(\tan(\theta) \sin(\phi))).
\end{align} 
Simplification of the $\delta$ function leads to
\begin{align} \label{eq:Abeltransform}
P_{\infty}(\psi)  =  \frac{1}{\pi \cos(\psi)}  \int^{\pi/2}_{0} \! d\theta  \, \frac{\cos (\theta ) P_{\infty}(\theta)}{\sqrt{\cos ^2(\psi )-\cos ^2(\theta )}}
\end{align} 
Combining the asymptotic behavior from Eq. (5) in the main text and \refn{eq:Abeltransform}, we obtain the following asymptotic behavior,
\begin{align} \label{eq:psiasymptotic}
P_{\infty}(\psi) \sim \psi ^{\frac{1}{B}-1} \frac{2^{-\frac{B+1}{B}}}{(B-1) B \sin \left(\frac{\pi }{2 B}\right)}, \quad \psi \ll 1,
\end{align}
which is valid for $B > 1$. From the latter \refn{eq:psiasymptotic}, we conclude that the velocity condensation is observable on the projected angle $\psi$ for a sufficiently strong magnetotactic constant. 
 
 \begin{figure}[h]
	\includegraphics[width=11cm]{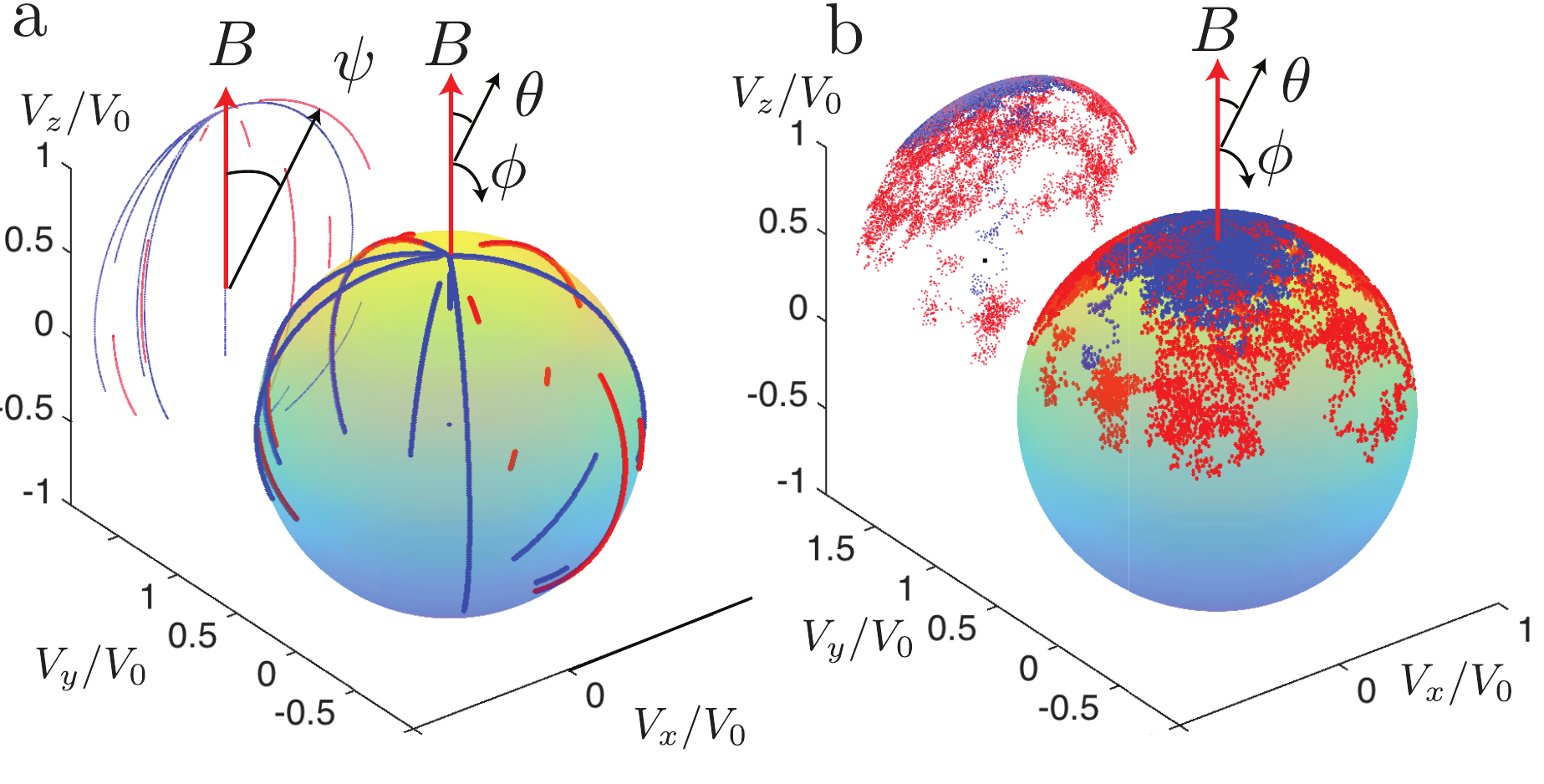}
	\caption{Trajectories of the velocity vector $V$, with a zenith angle $\theta$ and an azimuthal angle $\phi$, for (a) a RT walk and (b) an ABP; the magnetotactic constant is (blue line) $B=4$ and (red line) $B=1.8$. The angle $\psi$ is the projected alignment angle in the plane $y = 0$.}
	\label{fig:si:1}
\end{figure}

\begin{figure}[t!]
\includegraphics[width=11cm]{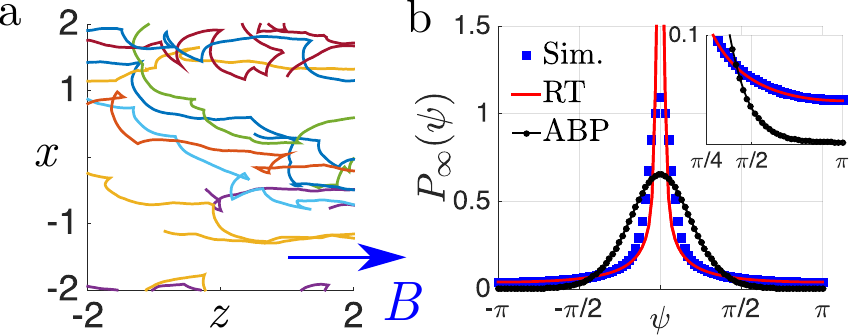}
	\caption{(color online) Simulations of tumbling  swimmers in the presence of a Brownian noise on the orientation ($D^{\prime}_r/\tau_r = 0.10$, $B = 3$). (a) Trajectories. Length scale is $V_0/\tau_r$. (b) Distribution of alignment angle with $B = 3$: (blue square) result of simulations at $t = 20 \tau_r$ after an initial tumble (red solid line), RT model with no rotary diffusion, and (black circle-solid line) best ABP fit, with $D_r/\tau_r = 0.60$. The maximal discrepancy with the cumulative distribution from simulation (i.e. Kolmogorov-Smirnov norm \cite{Feller1968}) is $3\%$ for the RT model, compared to $10\%$ for the ABP model.}
	\label{fig:3}
\end{figure}

\subsection{Approximate expressions for the velocity} \label{sm:sec:vel}

The averaged velocity in the direction of the magnetic field ($V_z = V_0 \left\langle \cos \theta \right\rangle$) is expressed in terms of the function $g_d(w)$, defined as:
\begin{align} \label{eq:gd}
 g_d(w) = \int_0^{\pi } \! d\theta \, {\cos (\theta ) \over \gamma_d \sin(\theta)} \left[{ 2 w \tan\left(\theta/2\right) \over (1 + w^2  
\tan\left(\theta/2\right)^2)}\right]^{d-1},
\end{align}
with $\gamma_2 = \pi$ and $\gamma_3 = 2$. We find that $g_2(w) = (w-1)/(w+1)$ in 2D and $g_3(w) = 2 \left(w^4-4 w^2 \log (w)-1\right)/\left(w^2-1\right)^2$ in 3D. We obtain an approximate expression for \refn{si:eq:vz} by expanding the function $g_d(w)$ around $w=1$. In 2D, we expand the function $g_2(w)$ at the $5$-th order around $w=1$, and we obtain:
\begin{align} \label{eq_v_approx_2D_1}
V_z/V_0  &\approx \frac{465 B^5+435 B^4+169 B^3+30 B^2+2 B}{4 (B+1) (2 B+1) (3 B+1) (4 B+1) (5 B+1)}. 
\end{align}
The expression of \refn{eq_v_approx_2D_1} is exact for $B \ll 1$ and reaches at $B \gg 1$ its maximal error, which is less than $4 \%$. \\

In 3D, we expand $g_3(w)$ around $w=1$ at the $2$-nd order, and we obtain:
\begin{align}  \label{eq_v_approx_3D}
V_z /V_0 &\approx  \frac{2 (B + 3 B^2)}{3 (1 + B) (1 + 2 B)}.
\end{align}
This expression \refn{eq_v_approx_3D} is exact for both $B \ll 1$ and $B \gg 1$ and it reaches at $B \approx 1$ its maximal error that is less than $5 \%$.
(maximal relative error is less than $1\%$).

\subsection{Diffusion coefficient} \label{sec:diffusion}

We now determine the diffusion coefficient $D_{\perp}$ in the transverse direction $(xOy)$, defined through the long-time limit of the transverse displacement $\left\langle\Delta y^2 \right\rangle \sim 2 D_{\perp} t$. We first notice that the variance of the transverse displacement after a single jump, denoted $\mathrm{Var}[Y_{1}]$, can be obtained after averaging the quantity $y(\theta_0, t) = \int^{t}_{0} \sin(\theta_{t}[\theta_0])$ over all $\theta_0$ and $t$. As successive runs are independent, this quantity is equal to the transverse diffusion coefficient: $D_{\perp}/D_0 =  \mathrm{Var}[Y_{1}]/\tau_r$.
In the large magnetic field limit $B \gg 1$, the transverse diffusion reads: $D_{\perp}/D_0 \underset{{B\gg 1}}{\sim}   {\gamma_d/B^2}$, where $\gamma_d = 3.3 \ldots$ in 2D and $\gamma_d = 0.46 \ldots$ in 3D. For ABPs, $D_{\perp}/D_0 \sim  {2/ B^2}$ for ${B\gg 1}$, as in this limit ABPs follow an Ornstein-Uhlenbeck dynamic with a $1/B$ spring constant (see \cite{Schienbein1993, Gardiner:2004}).  The fact that these transverse diffusion coefficients have the same scaling behavior suggest that, in spite of the velocity condensation, RTs are  as efficient as ABPs in exploring their environment.

\subsection{Regular runs} \label{sec:regular}

We now consider the case of runs of constant duration equal to $\tau_r = 1$ ($\rho(x) = \delta(x-1)$). This distribution is in particular used to model the motion of myxo-bacteria \cite{Balagam2015}. The distribution of spent time since the last tumble reads: $\pi(t) = \mathbb{1}_{0<t<1}$. From Eq. (2), we find that the probability distribution reads
\begin{align} \label{eq:P_sin_dir_2D}
P_{\infty}(\theta)  &=  \frac{1}{\pi B \sin \theta}  \left\lbrace 2 \arctan\left(e^{B} \tan \left(\frac{\theta }{2}\right)\right) - \theta\right\rbrace, \quad &\mathrm{(2D)}, \\
							&= -\frac{1}{2 B} \frac{1}{1/\tan(\theta )- \coth \left(B\right)/\sin(\theta )}, \quad &\mathrm{(3D)}.
\end{align}
Both expressions converge to a finite value at $\theta = 0$, equal to $\left\lbrace e^{B }-1\right\rbrace/(\pi B)$ in 2D, and $0$ in 3D.

The mean velocity reads $V_z/V_0 =  -2\log (2)/B+2\log \left(e^{B}+1\right)/B-1$ in 2D, and 
$V/V_0 = \coth \left(B\right)-1/B$ in 3D. Hence the expressions for the forward velocity in 3D is the same for the ABPs or RTs with regular steps, provided that identification  $D_r$ and $1/\tau_r$ are identified. This observation highlights the fact a Langevin fit of the velocity profile is not sufficient to discriminate between two microscopic models, namely the ABP model and the RT with regular steps.

\section{Pareto distribution} \label{si:sec:pareto}

For a Pareto distribution $\rho(x)= \mathbb{1}_{x>1} (\beta-1)/x^{\beta}$ with a finite mean run duration ($\beta >2$):
\begin{align} \label{eq:residualtime_par}
\pi(t)  = \mathbb{1}_{\left\lbrace 0 < t < 1 \right\rbrace}\frac{\beta-2}{\beta-1} +  \mathbb{1}_{ \left\lbrace 1<t \right\rbrace} \frac{\beta-2}{\beta-1} t^{1-\beta}.
\end{align}
The counter-intuitive effect known as the inspection paradox is that $\pi(t)$ has a broader distribution and fatter tails than $\rho(x)$ \cite{Feller1968}.  \\

\subsection{Probability density}

In order to use Eq. (2) from the main text, we notice that the function $t \rightarrow \sin(\theta_0^{\star})$ is sharply peaked around $t = -\log(\tan(\theta/2))/B$. By substitution of the expression $\sin(\theta_0^{\star})$ by a delta condition $\delta(t + \log(\tan(\theta/2))/B)$, we obtain the following approximate expression:
\begin{align}
P_{\infty}(\theta)\underset{\theta \rightarrow 0}{\sim} \gamma_d \frac{B^{\beta-1}  (\beta -2)}{ (\beta -1)} \frac{1}{\theta (\log (1/\theta ))^{\beta-1}}, 
\end{align}
with $ \gamma_d  = 4/(5 \pi) = 0.254$ (2D) and $ \gamma_d  = 8/25 = 0.32$  (3D), which are in close agreement to the value $\gamma  = 0.46$ obtained by fits to simulations, see Eq. (7).

\subsection{Averaged velocity}

In terms of the function $g_d(w)$ defined in \refn{eq:gd}, the averaged velocity reads
\begin{align} \label{si:eq:vz}
V_z/V_0 = \frac{\beta-2}{\beta-1}  \int_0^1 dt\, g_d(e^{-B t}) + \frac{\beta-2}{\beta-1}  \int_1^\infty dt\, t^{1-\beta} g_d(e^{-B t}).
\end{align}
We now determine the asymptotic behavior of $V_z/V_0$ at $B \ll 1$. In the limit $B \ll 1$,  we approximate $g_2(w) \approx (w-1)/2$ and $g_3(w) \approx 2(w-1)/3$, hence when $\beta \leq 3$:
\begin{align}
V_z/V_0 & =  \gamma_d B^{\beta-2} \frac{\Gamma (3-\beta )}{(\beta -1)},
\end{align}
where $\gamma_2 = 1/2$ and $\gamma_3 = 2/3$, see Eq. (8). The physical significance of the non-analytical $V_z/V_0 \propto B^{\beta-2}$ behavior is discussed in the main text. \\

\section{Non-magnetic torque functions}  \label{sec:ref}

In this section, we consider torques that take the general expression $f(\theta) = - \theta^n$ for all $\theta \in \left[0, \pi\right]$. We define the function $\theta_0 \rightarrow \theta-\theta_t[\theta_0]$, that is canceled for $\theta_0 = \theta_0^{*} = F^{(-1)}[F[\theta]-B t]$ -- where $F$ a primitive of $1/f$ and $F^{(-1)}$ is the reciprocal function of $F$. The function $t \rightarrow \theta^{*}_0(\theta, t)$ exceed the value $\pi$ for $t > t_c = (F(\pi) - F(\theta_0)/B$, i.e. for a run duration larger than $t_c(\pi)$, the point $\theta$ cannot be reached from an angle $\theta_0$ that is below $\pi$. 
With a modified torque
\begin{align} \label{eq:main_torque}
P_{\infty}(\theta) &= \frac{1}{f(\theta)} \int^{t_c(\theta)}_{0} \! \mathrm{d}t \, \pi(t)  (\mu. f)(\theta^{*}_0(\theta, t)).
\end{align}
If $F(\pi)$ is infinite, $t_c(\theta) = \infty$ and we retrieve the expression of Eq. (2).
 If $n < 1$, the $\theta = 0$ state can be attained in a finite time. Therefore, there is a finite probability that the particle reaches $\theta = 0$, the probability density has a an atom at $\theta = 0$. \\

\section{Perpendicular chemotaxis}  \label{sec:si_chemotaxis}
We now consider that the modulation of the run duration is given by $\tau_r(\theta) = 1 + \chi \sin(\theta)$. In this case, the Fokker-Planck equation reads:
\begin{align}  \label{eq:FPchemotaxis_perpendicular}
B \partial_\theta (\sin \theta P) - (1+\chi \sin \theta)  P(\theta) = -\left( 1 + \chi \left\langle P \sin \theta \right\rangle \right).
\end{align}
The solution to \refn{eq:FPchemotaxis_perpendicular} reads:
\begin{align}  \label{eq:solution_chemotaxis_perpendicular}
P(\theta)&= \gamma \exp(\chi \theta) \frac{\tan(\theta/2)^{\lambda_0}}{\sin(\theta)}  \smartit{\pi}{\theta}{\phi}  
\frac{\exp(-\chi \phi)}{\tan(\phi/2)^{\lambda_0}}.
\end{align}
As $\exp(\chi \theta)\rightarrow 1$ when $\theta \rightarrow 0$, we conclude that a finite value of $\chi > 0$ will not change of the critical magnetic field $B_c = 1$ obtained for $\chi = 0$.

\end{document}